\documentclass[preprint2]{aastex631}

\newcommand{\lya}{Ly$\alpha\,$}

\newcommand{\kms}{km s$^{-1}$}
\newcommand{\heii}{He II$\lambda$1640}
\newcommand{\TN}{TN~J1049-1258}
\usepackage[switch]{lineno}

\begin{document}

\title{An extended Lyman $\alpha$ outflow from a radio galaxy at z=3.7?}

\author[0009-0001-4945-5781]{Miguel Coloma Puga}
\affiliation{Department of Physics, Università di Torino \\
Via Pietro Giuria, 1, 10125, Torino, Italy}
\affiliation{INAF - Osservatorio Astrofisico di Torino \\
Via Osservatorio 20, I-10025 Pino Torinese, Italy}

\author[0000-0002-0690-0638]{Barbara Balmaverde}
\affiliation{INAF - Osservatorio Astrofisico di Torino \\
Via Osservatorio 20, I-10025 Pino Torinese, Italy}

\author[0000-0003-3684-4275]{Alessandro Capetti}
\affiliation{INAF - Osservatorio Astrofisico di Torino \\
Via Osservatorio 20, I-10025 Pino Torinese, Italy}

\author[0000-0002-1704-9850]{Francesco Massaro}
\affiliation{Department of Physics, Università di Torino \\
Via Pietro Giuria, 1, 10125, Torino, Italy}

\author[0000-0001-8353-649X]{Cristina Ramos Almeida}
\affiliation{Instituto de Astrofisica de Canarias\\
C. Vía Láctea, s/n, 38205 La Laguna, Tenerife, Spain}
\affiliation{Departamento de Astrofísica, Universidad de La Laguna \\
38206, La Laguna, Tenerife, Spain}

\author[0000-0003-2884-7214]{George Miley}
\affiliation{Leiden Observatory, Leiden University \\
PO Box 9513, 2300 RA Leiden, The Netherlands}

\author[0000-0001-8121-6177]{Roberto Gilli}
\affiliation{Osservatorio di Astrofisica e Scienza dello Spazio di Bologna\\
Via Gobetti 93/3, I-40129 Bologna, Italy}

\author[0000-0002-9889-4238]{Alessandro Marconi}
\affiliation{Dipartimento di Fisica e Astronomia, Università degli Studi di Firenze \\
Via G. Sansone 1,I-50019, Sesto Fiorentino, Firenze, Italy}
\affiliation{INAF - Osservatorio Astrofisico di Arcetri \\
Largo E. Fermi 5, I-50125, Firenze, Italy}

\begin{abstract}

Spatially resolved observations of AGN host galaxies undergoing feedback processes are one of the most relevant avenues through which galactic evolution can be studied, given the long lasting effects AGN feedback has on gas reservoirs, star formation, and AGN environments at all scales. Within this context we report results from VLT/MUSE integral field optical spectroscopy of TN J1049-1258, one of the most powerful radio sources known, at a redshift of 3.7. We detected extended ($\sim$ 18 kpc) \lya\ emission, spatially aligned with the radio axis, redshifted by 2250 $\pm$ 60 \kms\ with respect to the host galaxy systemic velocity, and co-spatial with UV continuum emission. This \lya\ emission could arise from a companion galaxy, although there are arguments against this interpretation. Alternatively, it might correspond to an outflow of ionized gas stemming from the radio galaxy. The outflow would be the highest redshift spatially resolved ionized outflow to date. The enormous amount of energy injected, however, appears to be unable to quench the host galaxy's prodigious star formation, occurring at a rate of $\sim$4500 M$_{\odot} yr^{-1}$, estimated using its far infra-red luminosity. Within the field we also found two companion galaxies at projected distances of $\sim$25 kpc and $\sim$60 kpc from the host, which suggests the host galaxy is harbored within a protocluster.

\end{abstract}

\section{Introduction} \label{sec:intro}

Outflows are believed to be the main driver through which active galactic nuclei (AGN) regulate the evolution of their host galaxies and affect the large scale environment (AGN feedback), operating mostly at the epoch of the peak of star formation between redshift 2 and 3 \citep{hopkins2006star,Miley_2008}. Outflows, extended on kiloparsec scales, powered by the AGN radiative pressure (quasar feedback mode) and/or radio jets (radio feedback mode), have been resolved in ionised, atomic, and molecular gas both in local (e.g. \citealt{Rupke_2011}, \citealt{Rupke_2013}, \citealt{Cicone_2012}, \citealt{Rodr_guez_Zaur_n_2013}, \citealt{Speranza_2021}, \citealt{Ramos_2022}) and in high redshift (radio loud and radio quiet) quasars (e.g. \citealt{Alexander_2010}; \citealt{Nesvadba_2010}; \citealt{Maiolino_2012}; \citealt{Cano_D_az_2012}; \citealt{Cresci_2015}). Characterizing and constraining the effects and extent to which AGN feedback affects star formation is one of the most important problems in modern astrophysics, as it has ties to the biggest questions in galaxy evolution and cosmology.

In this context, high redshift radio galaxies (HzRGs) are particularly relevant as the nuclear regions are obscured along our line of sight due to a circumnuclear dusty structure (e.g., \citealt{antonucci93}) or by the host galaxy itself. The lower nuclear continuum luminosities allow us to reduce the observational uncertainties produced by the subtraction of a prominent Point Spread Function (PSF), probing the host and the gas structure down to the limit of the spatial resolution of only a few kpc. 

We started a comprehensive program of observations with the Multi Unit Spectroscopic Explorer (MUSE) at the VLT to explore the properties of the ionized gas in HzRGs. In this Letter we present the results obtained for TN J1049-1258, an HzRG at redshift 3.697 $\pm$ 0.004 \citep{Bornancini_2007}, located at $\alpha$=10:49:06.2, $\delta=$-12:58:19 (J2000). The scale factor at this redshift (assuming an H$_{0}$=69.6 and $\Omega_m$=0.286 cosmology) is 7.3 kpc arcsec$^{-1}. $ Its radio flux measured at 74 MHz by the NRAO VLA Sky Survey is 5.18 $\pm$ 0.56 Jy while its radio spectral index is $\alpha = 1.4$ \citep{Condon_1998}.\footnote{Spectral index, $\alpha$, is defined by flux density, S$_{\nu}\propto\nu^{-\alpha}$.} The jet power, $P_{\rm jet}$, can be obtained using the correlation between $P_{\rm jet}$ and radio luminosity, based on the measurement of the mechanical power of radio sources producing cavities in the surrounding hot gas as proposed by \citep{Cavagnolo10}. The radio luminosity of TN J1049-1258 at (rest frame) 327 MHz can be estimated as $P_{327} = 2.7\times10^{44}$ erg s$^{-1}$ which yields a jet power of $P_{\rm jet} \sim 6 \times 10^{45}$ erg s$^{-1}$. A similar value is obtained by using the relation between radio and jet power from \citet{willott99}.
In the Karl G. Jansky VLA Sky Survey (VLASS, \citealt{Lacy_2020}), performed at 3 GHz with a resolution of 2\farcs5, its radio emission is dominated by two radio components separated by 11\arcsec\ (80 kpc) oriented at the position angle -80$^\circ$, with the host located close to their midpoint. Its radio luminosity at the rest frame frequency of 500 MHz is 8.7$\times10^{28}$ W Hz$^{-1}$, making TN J1049-1258 one of the most luminous radio sources known (see \citealt{Miley_2008}). The flux at 3.5 $\mu$m, corresponding to the optical emission in the rest frame, measured by Wide-field Infrared Survey Explorer (WISE, \citealt{Wright_2010}) of TN J1049-1258 indicates a luminosity of $2\times10^{11} L_\odot$, suggesting that its host is already a very massive galaxy. 

\section{Observations and data analysis}\label{sec:obs}

The observations were carried out with the Multi Unit Spectroscopic Explorer (MUSE) on the 29th of January of 2022 as part of the program ID:108.22FU. Four separate observations were performed, between which the telescope was rotated 90$^\circ$ to reject cosmic rays, for a total of 2800s of exposure time. The seeing measured from several point sources in the field of view was estimated as 0\farcs71$\pm$0\farcs04. We used the ESO MUSE pipeline (version 2.8.7) to obtain a fully reduced and calibrated data cube \citep{Weilbacher20}.
We derived an absolute astrometric calibration of the MUSE data by cross-matching visible sources in both MUSE and Pan-STARRS (Panoramic Survey and Rapid Response System \citep{Chambers_2016}). Based on the scatter of the sources considered we estimate the derived astrometry to have an error of $\sim0\farcs2$.

\begin{figure}[h]
    \centering
    \includegraphics[width=0.5\textwidth]{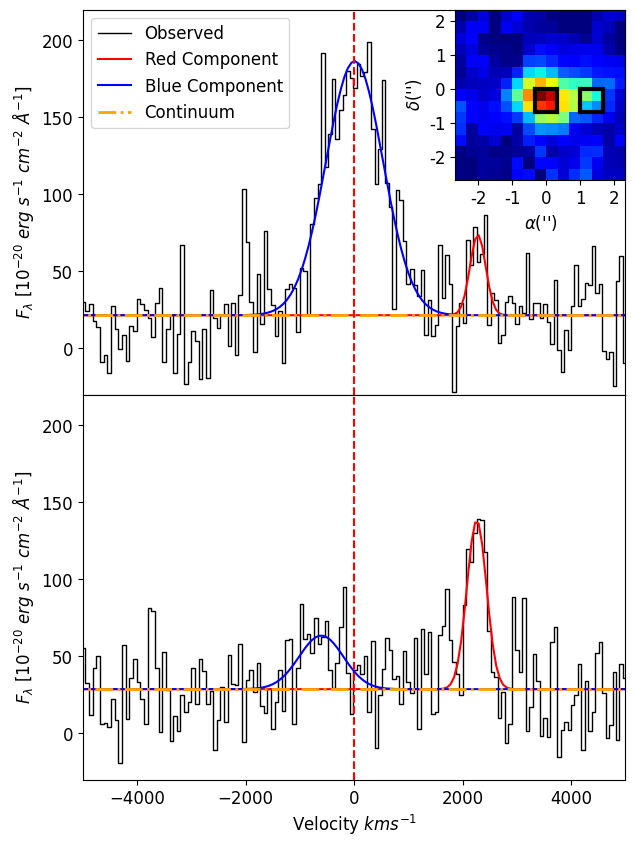}
    \caption{\textbf{Top}: Nuclear spectrum of TN J1049-1258, extracted from a 0$\farcs$8 $\times$ 0$\farcs$8 synthetic aperture, along with the modelled continuum and fitted lines. The spectrum shows the presence of two components, separated by $\sim 2,250$ km s$^{-1}$. In the top right inset we show the total \lya\ flux map, spatially rebinned into 2$\times$2 pixels, with the two synthetic apertures used to extract the nuclear and off-nuclear spectra superimposed. \textbf{Bottom}: Off-nuclear spectrum extracted 1$\farcs$6 west of the nucleus. The redshifted line becomes the brightest component. In both cases, the red dashed line marks the host galaxy's systemic velocity, corresponding to a redshift of 3.697. The X and Y scales are identical in both spectra.}
    \label{fig:centralspec}
\end{figure}

After first inspecting the nuclear spectrum on a large aperture around the peak of \lya\ emission, we found two emission lines separated by about 50 \r{A}. To fully characterize both the emission line morphology and the gas kinematics, we then performed a two Gaussian component fit in each spaxel (see Fig. \ref{fig:centralspec} for the nuclear and off-nuclear spectra). Aiming to increase the signal to noise ratio per pixel, the cube was spatially rebinned into 2$\times$2 pixels, resulting in a 0\farcs4$\times$0\farcs4 area.  The \textit{modeling} tool from \textit{Astropy}'s Python package (\citealt{2013A&A...558A..33A}, \citealt{2018AJ....156..123A}, \citealt{The_Astropy_Collaboration_2022}) was used to model and fit both the continuum and emission lines. For the continuum we used a linear least squares fit, while the emission lines were modelled with a Gaussian using a Levenberg-Marquardt least squares fit. 

From the fit we obtained the distribution of total flux and first and second moments for both components, while the uncertainties of the parameters were obtained through a Monte Carlo simulation. Each realization consists of adding flux variations to each spectral pixel, with the value of the variations taken at random from a normal distribution centered at the pixel noise. From each realization, a value for the relevant parameters (line center, amplitude, width) is obtained. After a set number of iterations (100 in our case) we obtain a value distribution for each of these parameters, from which we extract the mean (parameter value) and 1$\sigma$ standard deviation (parameter uncertainty). The pixel noise was calculated as the 1$\sigma$ standard deviation of the flux in a 50 nm interval centered on \lya\ from a section of empty sky of equal size as the cube extracted for the line analysis (80$\times$80 pixels or 16\arcsec$\times$16\arcsec). All emission line maps have been obtained by setting a S/N threshold at 3.

\begin{figure}[htbp]
    \centering
    \includegraphics[width=0.47\textwidth]{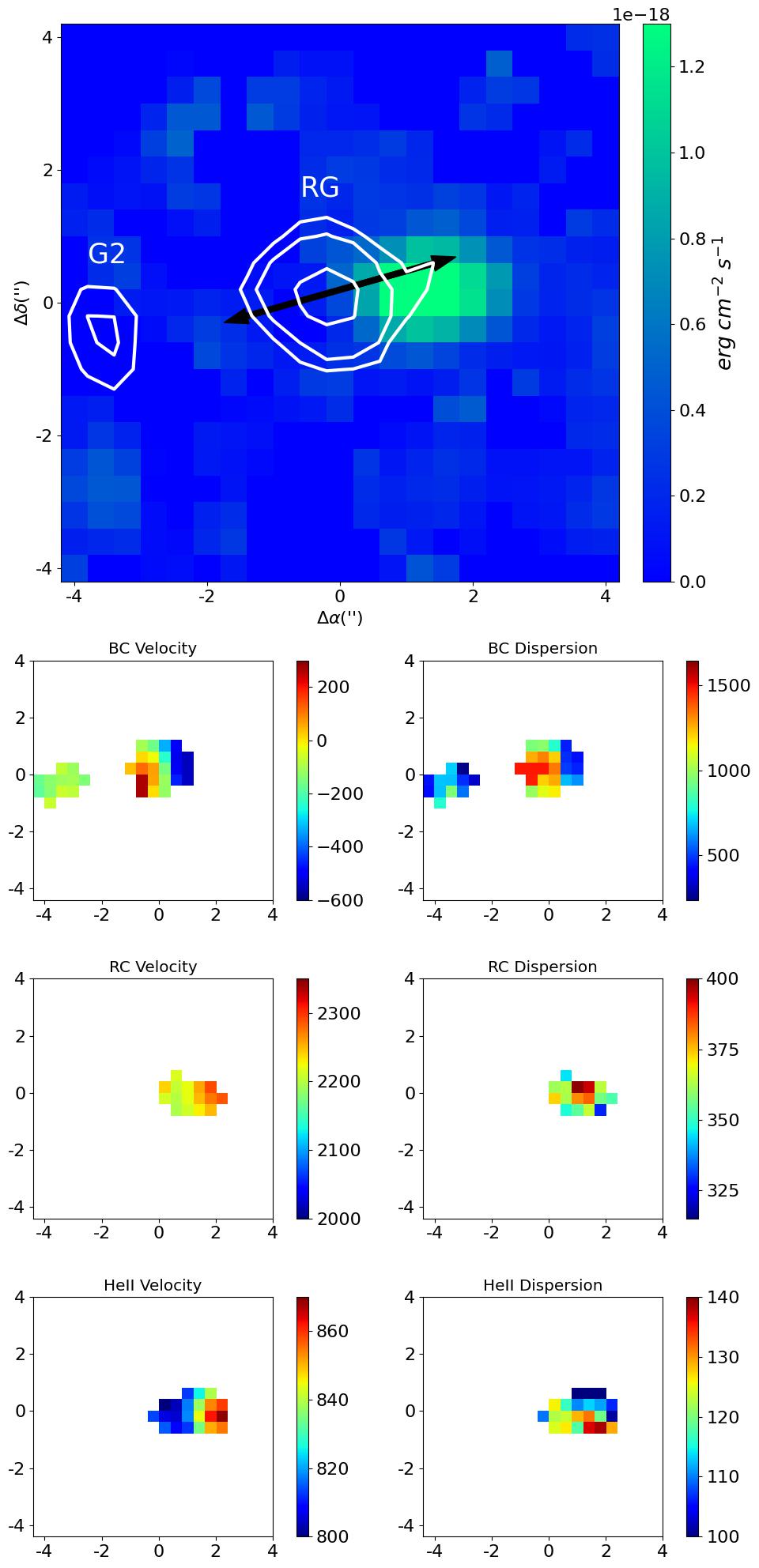}
    \caption{\textbf{Top}: \lya\ emission map of the red component, with superimposed iso-intensity contours of the \lya\ from the host galaxy. The field of view is $\sim 30 $ kpc $\times 30$ kpc. There is a clear offset between the center of the host galaxy emission and of the red component. A second source of \lya\ emission (that we name G2) is seen $\sim$ 3.5\arcsec\ east of the host galaxy. The arrow shows the radio axis. \textbf{Bottom}: maps of projected velocity and velocity dispersion of the host galaxy (top), of the red \lya\ component (middle), and of the HeII line in \kms. In all cases, the S/N threshold on the line fit is set at 3$\sigma$.}
    \label{fig:velmaps}
\end{figure}

The continuum map was obtained using multiple intervals within MUSE's wavelength range (475 nm to 935 nm, 101 nm to 198 nm rest frame) over a total range of 144 nm, selected to avoid both sky lines as well as object emission lines.

\section{Results} \label{sec:res}

We estimated the redshift of \TN\ from the fit of the main component of \lya\ line in the nuclear spectrum, which corresponded to z=3.697$\pm$0.001, the same value obtained by \citet{Bornancini_2007}. However, it must be noted that the \lya\ line peak has been observed to be redshifted from the systemic velocity (measured by the H$\alpha$ emission) by $\sim$450 \kms\ on average, and by as much as $\sim 900$ \kms\ (e.g., \citealt{steidel2010}).

The emission maps of both components and all relevant velocity and velocity dispersion maps are shown in Fig. \ref{fig:velmaps}. The velocity field estimated from the host galaxy's \lya\ emission presents a large gradient, with a velocity range of $\sim$900 \kms\ and a velocity dispersion ranging from 500 to 1500 \kms, that can be ascribed to either rotation or radial motion. While there is a clear spatial correlation between width and relative velocity, interpreting the absolute values of the velocity dispersion is made difficult by the resonant character of \lya\ emission. Neutral hydrogen in the line of sight of the emitting gas will result in peaks and troughs, deeply modifying the original line profile depending on the density, distribution, and velocity of the neutral H gas.

The redshifted component, has an elongated structure extending $\sim$2.5\arcsec, i.e., $\sim$ 18.5 kpc, measured as the distance between the two farthest pixels with S/N$>$3, along the same position angle of the radio structure. Its relative velocity, measured using the line profile in the off-nuclear spectrum (see Figure \ref{fig:centralspec}) is $v=2250 \pm 60$ \kms\ and it remains remarkably similar throughout the spatial extent of the emission zone. In a similar fashion, the velocity dispersion values are within a range of $\sim$70 \kms, with this red component being much narrower than the one in the host galaxy; its FWHM measured from the off-nuclear spectrum is $\sigma_{\rm RC}=430\pm20$ \kms.

\begin{figure}[ht]
    \centering
    \includegraphics[width=0.5\textwidth]{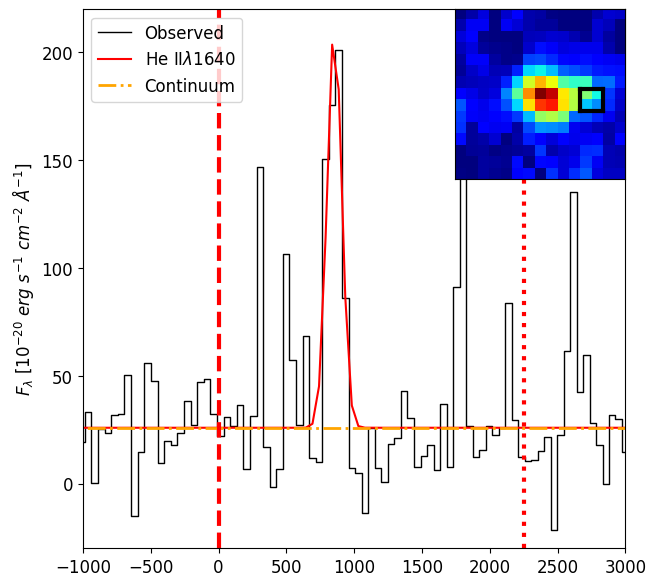}
    \caption{He II spectrum extracted using the same off-nuclear aperture as \lya. The dashed line marks the host systemic velocity, and the dotted line marks the extended emission velocity.}
    \label{fig:heii}
\end{figure}

In the region cospatial to the \lya\ red component we also detected emission from the HeII$\lambda$1640 line (see Fig. \ref{fig:heii}). This line is blueshifted by 1404$\pm$9 \kms\ (value obtained from the off-nuclear spectrum shown in Fig \ref{fig:centralspec}) with respect to the \lya\ emission. Its velocity and velocity dispersion maps are shown in Figure \ref{fig:velmaps}. From this spectrum we obtain a line ratio Ly$\alpha$/He~II=$1.5\pm0.3$, which is at the very low end of the values measured in HzRGs (see \citealt{Villar_Mart_n_2007}).

\begin{figure}
    \centering
    \includegraphics[width=0.5\textwidth]{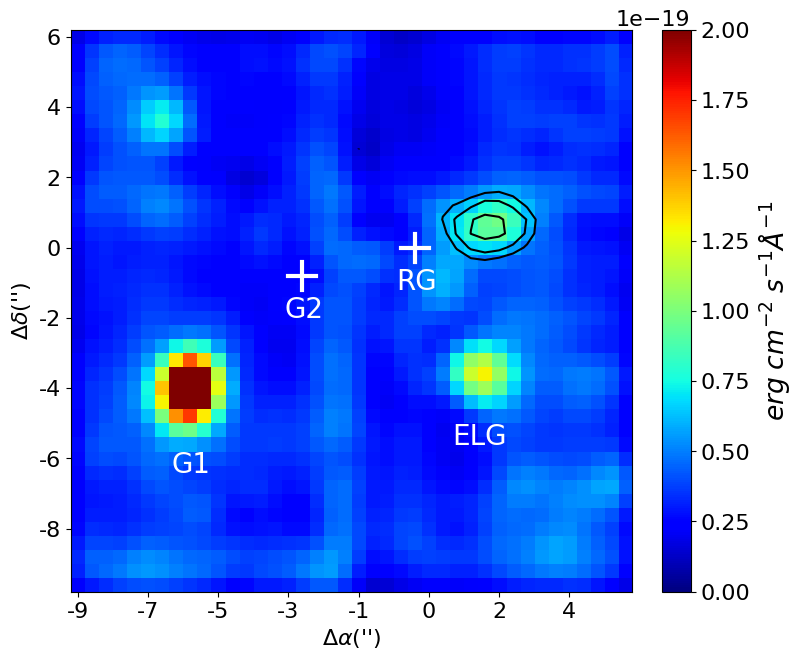}
    \caption{Rest frame UV continuum flux density map of the field around  TN~J1049-1258. A source of continuum can be seen to the west of the \lya\ emission center, cospatial with the red component (black contours at 0.5, 0.8 and 1.3 $10^{-18}\;erg\;cm^{-2}\;s^{-1}$). The brightest UV source, G1, is located $\sim 8$\arcsec\ ($\sim 60$ kpc) southeast of the host galaxy. At the location of G2 (marked with a "+" sign) an image defect (the vertical line at $\Delta\alpha\sim$-1'') does not allow us to obtain a robust UV counterpart to the \lya\ emission seen in Fig.\ref{fig:velmaps}. Another source found 5\arcsec\ to the south and slightly west is identified as a z=0.87 emission line galaxy and labelled ELG}.
    \label{fig:continuum}
\end{figure}

\begin{figure}
    \centering
    \includegraphics[width=0.5\textwidth]{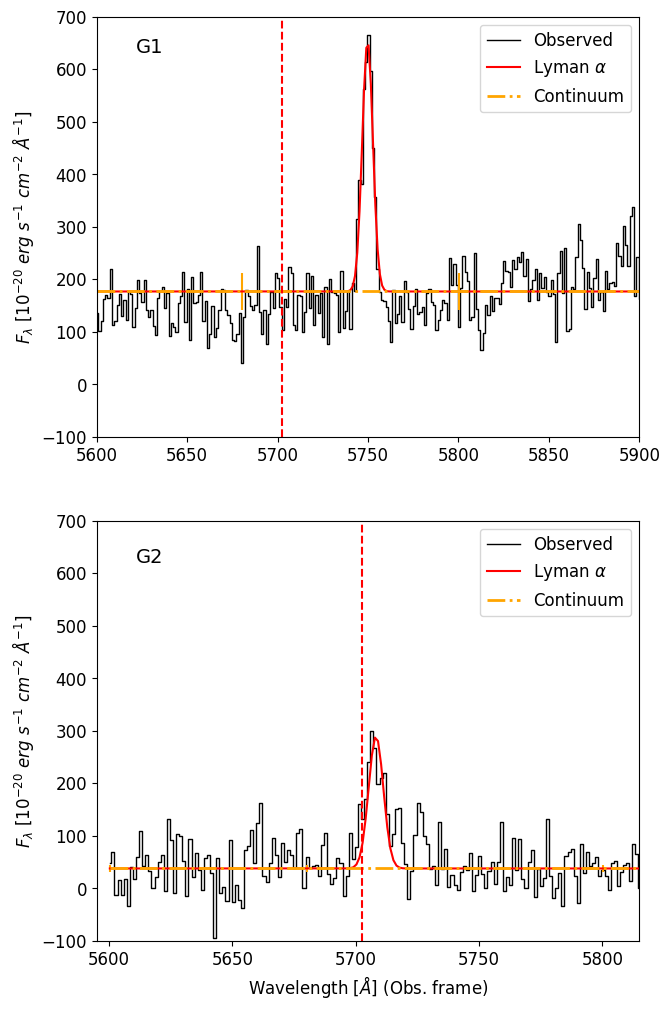}
    \caption{Spectra in the \lya\ region for the two companion galaxies, G1 (top) and G2 (bottom). The red dashed line marks the RG's redshift measured with \lya.}
    \label{fig:compspec}
\end{figure}

UV continuum emission, within the rest frame range $\sim 1,000-2,000$ \AA, (see Fig. \ref{fig:continuum}) is observed on the west side of the host galaxy, cospatial with the \lya\ red component. The source observed $\sim$5\arcsec\ south of the radio galaxy (RG from hereon) is a galaxy at z=0.87, based on the identification of the [O~II]$\lambda\lambda$3726,3729 doublet as well as H$\delta$ and H$\gamma$ in its spectrum. The brightest UV source in the field is another potential companion we named G1, located $\sim$8$\arcsec$ ($\sim$60 kpc) southeast of the host galaxy. Its \lya\ emission was detected at a very similar redshift (see Fig. \ref{fig:compspec}, top panel) and its high redshift identification is also supported by it being a g-dropout based on the photometry from Pan-STARRS \citep{Chambers_2016}.
\footnote{The Pan-STARRS aperture magnitudes for G1 are g=24.08$\pm$0.02, r=22.44$\pm$0.01, and i=22.39$\pm$0.01, respectively, then satisfying the g-dropout criteria by \citet{pouliasis22}.}

To the east of the radio galaxy, another patch of \lya\ emission can be seen in the top panel of Fig. \ref{fig:velmaps}. This is associated with a companion galaxy we named G2, whose identification is possible thanks to the presence of \lya\ (see Fig. \ref{fig:compspec}, bottom panel), \heii, and CIV$\lambda\lambda1548,1550$ in its spectrum. Its redshift, z=3.685$\pm$0.002, was estimated from the \heii\ emission, which is a non resonant line. The presence of G1 and G2 suggests TN~J1049-1258 may be harboured within a proto-cluster.

The main result of these observations is the presence of extended \lya\ emission redshifted by $\sim$2,250 \kms\ and we can envisage two possibilities for its origin: 

     \textbf{1) A companion galaxy.} The presence of UV continuum emission co-spatial with the \lya\ region suggests that this might be associated with another galaxy at a similar redshift. This interpretation is also supported by the low velocity dispersion ($\sigma=430\pm 20$ \kms) of the \lya\ emission. The \lya\ equivalent width (EW) of $6.6 \pm 2.0$ \AA\ does not provide useful information as to the origin of the line emission since the EW of Lyman Alpha Emitters spans a considerably wide range, with measurements ranging from a few \AA\ to several hundreds (see \citealt{Kerutt22}). However, studies of high redshift galaxies found that the typical radius of a z $\sim$4 galaxy is 0\farcs2, that is, 1.5 kpc \citep{Ferguson04,Bouwens04}. The full width at half maximum of the \lya\ red component emission is significantly larger: by fitting its spatial distribution with a Gaussian we obtain a size of 1\farcs79$\pm$0\farcs14, equivalent to a deconvolved extension of R=12.0$\pm$0.9 kpc. We must note, however, that G1 is also extended, with a deconvolved size of 7.3$\pm$0.5 kpc. The strongest argument against this interpretation is the blueshift of $\sim$1,400 \kms\ which the \heii\ line presents with respect to the extended \lya\ line, as it greatly exceeds the range of velocity offsets between \lya\ and nebular lines observed in high redshift sources \citep{steidel2010}, arguing in favour of a different origin.
     
    \textbf{2) An outflow of ionized gas.}  This hypothesis is supported by simulations of the theoretical profile of \lya\ \citep{Ahn_2004,Verhamme_2006,Behrens_2014}. In particular, the models by \citet{Behrens_2014} show that a bipolar expanding shell, a reasonable approximation for a nuclear outflow, can produce a secondary red-shifted peak. This unusual line profile is explained by the fact that \lya\ is a resonant emission line, subject to the effects of radiative transfer due to the large cross-section of interaction between \lya\ photons and neutral hydrogen atoms. As a consequence, the ubiquity of neutral hydrogen not only within galaxies but in the intergalactic medium along our line of sight prevents most, if not all, light blueward of the central \lya\ wavelength from emerging and, as is the case here, only the red side is observed. These effects can also explain the large velocity difference between the \lya\ and the He~II.
    
     Additionally, the presence of UV continuum aligned with the radio axis  is a common feature of HzRGs and has been explained as due to a combination of scattered nuclear light, fast shocks, and jet-induced star formation (see, e.g., \citealt{McCarthy93} or \citealt{Miley_2008}), which also points to an in-host origin for the emission.

With all of these considerations taken into account, we conclude that the outflow hypothesis is the more plausible scenario.

We estimated the mass of the outflow, $M_{out}$, from the H$\beta$ luminosity, $L_{H\beta}$, using the relation derived by \citet{osterbrock89}:

\begin{equation}
    M_{out}=7.5\,\times 10^{-3} \left(\frac{10^{4}}{n_e}\frac{L_{H\beta}}{L_\odot} \right) M_\odot
\end{equation} assuming the ratios Ly$\alpha/$H$\alpha$=8.7 
\citep{Sobral_2019} and H$\alpha$/H$\beta$=2.86, and a gas 
density of n$_e$ = 200 cm$^{-3}$, which enables a comparison between the energetics of this object and a sample of low redshift radio loud AGN that utilizes this same value \citep{Speranza_2022}.

We derived the following physical quantities (see e.g. \citealt{Fiore_2017}) for the outflow:
\begin{itemize}
    \item mass outflow rate:  $\dot{M}=3v_{out}\frac{M_{H\beta}}{R_{out}}= 3.4 M_\odot$ yr$^{-1}$
    \item kinetic energy: E$_{kin}=\frac{1}{2}M_{H\beta}v_{out}^{2}=10^{56.6}$ erg
    \item kinetic power  $\dot{E}_{kin}=\frac{1}{2}\dot{M}v_{out}^{2}= 10^{42.4}$ erg s$^{-1}$
\end{itemize} 

where $v_{out}=2,250$ \kms\ and $R_{out}=18.5$ kpc are the outflow velocity and full extension respectively.

These estimates should be considered as lower limits for two reasons: first of all we cannot correct for the effects of projection and, in addition, there might be a contribution of collisional excitation to the production of emission lines.

\section{Discussion and conclusions}\label{sec:disc}

The star formation rate (SFR) can be estimated from the ultraviolet (UV) and far infrared (FIR) emission of the source. While the UV light directly traces the young stellar population, the FIR emission is produced by warm dust heated by the ultraviolet photons emitted by young stars. We found the UV contribution to the SFR estimate for the RG to be negligible; it's also worth mentioning that UV continuum is subject to various effects that make it a more unreliable tracer at higher redshift \citep{Wilkins_2012}. 

Conversely, TN~J1049-1258 has been detected by pointed far infrared observations from the Spectral and Photometric Imaging Receiver (SPIRE), onboard the Herschel satellite \citep{Poglitsch_2010,Griffin_2010}. At 250 $\mu$m and 350 $\mu$m, the observed fluxes are 63.0$\pm9.5$ mJy and 67.7$\pm11.8$ mJy, respectively. These frequencies correspond to $\sim$ 50 and 70 $\mu$m in rest frame, which coincides with the peak of the warm dust emission, heated by star forming regions. Measurements from WISE were obtained on the W1 and W2 passbands (W3 and W4 were upper limits). From a fit to the spectral energy distribution (SED) using the WISE and Herschel measurements (see \citealt{Balmaverde16} for details) the derived total FIR luminosity is L$_{(8-1000 \mu m)}$ = 3.2$\times 10^{13} L_\odot$, which makes this source a hyper luminous infrared galaxy. It corresponds to a prodigious star formation rate of SFR$\sim$4,500 M$_{\odot}$ yr$^{-1}$, among the highest values ever estimated (see the compilation of SFRs from \citealt{Lagache18}). However, due the low spatial resolving power of Herschel (the Herschel data reduction guide suggested to adopt an aperture photometry radius for the 250 $\mu$m and 350 $\mu$m images of 22\arcsec and 30\arcsec, respectively), both companions and the z=0.87 galaxy are included within the aperture and thus there may be contamination from these sources. Additionally, there might be an AGN contribution. However, the torus emission peaks at 20 $\mu$m to 30 $\mu$m \citep{2008ApJ...685..160N} and in order to reproduce the FIR fluxes this component would exceed the WISE measurements. We conclude that the AGN can only be a minor contributor to the Herschel fluxes.

Regarding the energetics of the system, the jet power is $\sim$4 orders of magnitude larger than the outflow kinetic power. \citet{Speranza_2021} estimated the outflow kinetic power of a sample of radio galaxies at low redshift: while they measured values of $\dot{E_{kin}}$ similar to that of TN~J1049-1258, they found significantly larger values of $\dot{E_{kin}}/P_{\rm jet}$, typically $\sim 0.1$. This suggests that the relativistic jets and the outflow may be decoupled and produced by two independent acceleration mechanisms: the jets might be driven by the magnetic field in the innermost regions of the accretion disk while the outflow is powered by radiation pressure. Alternatively, the ionized gas traced by the \lya\ emission might just be the tip of the iceberg of a larger amount of outflowing gas, because this line is a poor tracer in terms of representative masses and luminosities. Dedicated radiative transfer simulations with geometries and parameters resembling those observed in TN 1049-1258 will be essential to fully model the results of these observations. To obtain a comprehensive view of the energetic of the outflow, we should also consider other ionized gas tracers, e.g. the optical [O~III] line, unaffected by resonance and absorption effects. Moreover, a relevant contribution to the energetic of the outflow is expected to originate from the molecular gas component, which can be studied using lines produced by the CO molecule transitions or the [C~II]158$\mu$m emission.

Another source of uncertainty in the outflow estimates are the ionization mechanisms in play, as collisional excitation may contribute. In fact, the low Ly$\alpha$/He~II ratio observed in the extended emission is not dissimilar to the one obtained in \citealt{Scarlata_2009} where, within another high redshift (z=2.38) system of Lyman Alpha Emitters, one of the sources displayed a comparably low ratio (Ly $\alpha$/He II=2.2$\pm$1.0) and the mechanism of origin was suspected to be accretion of cold gas onto the dark matter halo, resulting in collisional excitation.

Despite the large amount of energy injected into the host galaxy by the ionized outflow and the relativistic jets, TN~J1049-1258 displays globally a very large star formation rate, although the companion galaxies might also contribute to the Herschel-measured flux. Rest frame FIR observations at high spatial resolution, e.g., those that can be produced by facilities such as the Atacama Large Millimeter/sub-millimeter Array (ALMA), will be fundamental to obtain a more accurate estimate of the SFR in the host galaxy, and if possible, resolve its spatial distribution in a bid to better understand the interplay between jets, outflows, and gas reservoirs.

The possibility of outflows and more generally, jets, inducing star formation has been studied in both galactic (see \citealt{Bicknell_2000,Miley_2008,Capetti_2022}) and cosmological scales (\citealt{Gilli_2019}), as well as simulations (\citealt{Gaibler_2012}). As such, it is possible that the observed star formation from the RG is either induced or enhanced by the jet. Spatially resolved observations of cold molecular gas could answer questions such as where the star formation is occurring: whether jet-driven shocks are pressurizing gas bubbles in the intergalactic medium and causing fragmentation, or the expanding cocoon of the jet is inducing similar phenomena within the host galaxy itself. The co-spatiality between UV continuum and outflow suggests the former, though a mixture of both is not out of the question.

The presence of TN J1049-1258 inside a protocluster is within expectations. Studies have found that up to 75\% of powerful HzRGs reside within these structures \citep{Venemans_2006}. This is understandable considering that the powerful AGN responsible for this radio emission require massive galaxies, which in turn inhabit rich, dense environments that act as fertile soil for cluster-like structures \citep{Hatch_2014}. 
There is a great deal of relevance in protoclusters within the $\Lambda$CDM model, as they represent massive structures hypothetically linked by large dark matter halos: understanding how high-z and nearby galaxy clusters relate to each other is essential in unveiling the behaviour and characteristics of dark matter. Their evolution through the cosmic ages as well as their relationship to powerful RLAGN which likely become the brightest cluster galaxies are ongoing and widely studied topics within cosmology and high-redshift astrophysics.


\vspace{5mm}
\facilities{VLT(MUSE), Herschel, VLA, WISE}

\bibliography{sample631}{}
\bibliographystyle{aasjournal}

\clearpage
\newpage

\end{document}